\documentclass[conference]{IEEEtran}
\IEEEoverridecommandlockouts
\usepackage{cite}
\usepackage{amsmath,amssymb,amsfonts}
\usepackage{algorithmic}
\usepackage{graphicx}
\usepackage{textcomp}
\usepackage{xcolor}
\def\BibTeX{{\rm B\kern-.05em{\sc i\kern-.025em b}\kern-.08em
    T\kern-.1667em\lower.7ex\hbox{E}\kern-.125emX}}
\begin{document}
\setlength{\textfloatsep}{5pt}
\title{Unsupervised Machine Learning Identifies Latent Ultradian States in Multi-Modal Wearable Sensor Signals\\
\thanks{This work is supported by a UKRI Future Leaders Fellowship held by Y.W. (MR/V026569/1) and a UKRI EPSRC CLOSE-NIT Network Plus grant held by Y.W and C.T.}
}

\author{\IEEEauthorblockN{Christopher Thornton}
\IEEEauthorblockA{CNNP Lab\\ School of Computing\\ Newcastle University\\ United Kingdom\\ email:chris.thornton@ncl.ac.uk}
\and
\IEEEauthorblockN{Billy C. Smith}
\IEEEauthorblockA{CNNP Lab\\ School of Computing\\ Newcastle University\\ United Kingdom}
\and
\IEEEauthorblockN{Guillermo M. Besn\'e}
\IEEEauthorblockA{CNNP Lab\\ School of Computing\\ Newcastle University\\ United Kingdom}
\and
\IEEEauthorblockN{Bethany Little}
\IEEEauthorblockA{CNNP Lab\\ School of Computing\\ Newcastle University\\ United Kingdom}
\and
\IEEEauthorblockN{Yujiang Wang}
\IEEEauthorblockA{CNNP Lab\\ School of Computing\\ Newcastle University\\ United Kingdom}}

\maketitle

\begin{abstract}
Wearable sensors such as smartwatches have become ubiquitous in recent years, allowing the easy and continual measurement of physiological parameters such as heart rate, physical activity, body temperature, and blood glucose in an every-day setting. This multi-modal data offers the potential to identify latent states occurring across physiological measures, which may represent important bio-behavioural states that could not be observed in any single measure. Here we present an approach, utilising a hidden semi-Markov model, to identify such states in data collected using a smartwatch, electrocardiogram, and blood glucose monitor, over two weeks from a sample of 9 participants. We found 26 latent ultradian states across the sample, with many occurring at particular times of day. Here we describe some of these, as well as their association with subjective mood and time use diaries. These methods provide a novel avenue for developing insights into the physiology of everyday life. 
\end{abstract}


\section{Introduction}
Wearable sensors such as smartwatches, glucose monitors, and portable electrocardiograms, offer the potential to measure physiological processes throughout the body in an everyday setting. These measures can reflect a combination of intrinsic biological rhythms \cite{Upasham2021,Meyer2022}, the behaviours of the wearer \cite{Constantino, Kolehmainen2023, Stavropoulos2021, Perski2022}, and responses to external events that may stress or arouse the wearer \cite{Zhang2021, Sano2018}. Teasing out these contributions is an ongoing challenge for those wishing to use wearable sensor derived physiological measures to indicate emotional state \cite{Schmidt2019, Morales2018,Gedam2021}, external events \cite{Kyriakou2019}, or intrinsic biological rhythms \cite{Cui2023}.

An additional challenge is that many physiological states may not be apparent when observing a single measure, but only become apparent when considering multiple measures at once. For example increased blood glucose in response to eating a meal is distinct from increased blood glucose in response to exercise \cite{Sevil2021}. This distinction can only be made by combining information from multiple measures (e.g. blood glucose concentration and movement). Similarly, the additional context provided by multiple  modalities is important for the interpretation of measures of the autonomic nervous system such as electrodermal activity and heart rate variability \cite{Ghiasi2020}. 

Many physiological changes occur on timescales ranging from 30 minutes to 12 hours - often referred to as episodic ultradian events or fluctuations \cite{Goh2019}. These can be driven by behaviour or environment such as when engaging in exercise, eating, sleeping, or being stressed, but also can be driven by endogenous rhythms or events such as the pulsatile secretion of cortisol \cite{Russell2019} or ultradian sleep cycles \cite{Cajochen2024}. Quantifying the time spent in these ultradian states, including in latent states, is an important step towards understanding individual differences in physiology and health behaviours. 

Hidden Markov models (HMM) are often used when one has an observable series of measurements $Y_t = (Y_1 ... Y_N)$ and from this wishes to estimate a process $C_t = (C_1 ... C_N)$ which has not been observed. When the HMM is used in an unsupervised manner, the hidden states of process $C_t$ may be latent states that can be interpreted only through the distributions of observed measurements associated with them \cite{Suarez2021}. This is ideal when we do not have a set of specific states we wish to detect, but rather wish to determine the physiological states which emerge naturally from the data.

One limitation of the HMM is that the time spent in each state must follow a geometric distribution \cite{Langrock2011}. The hidden semi-Markov model (HSMM) is an extension of the hidden Markov model (HMM) where the duration of each hidden state in the unobserved process $C$ is captured explicitly by modelling a duration distribution allowing each state to have an expected duration \cite{Bulla2006} - as would be expected for behavioural and physiological states \cite{Suarez2021}. The HSMM has been used in an unsupervised manner previously to segment and cluster single channel wearable sensor data such as accelerometry \cite{Thornton2023, VanKuppevelt2019}, and to identify latent states in financial markets \cite{Yao2020}.

Here we aim to extend this data-driven approach to multiple channels of data, and to use it to identify states that exist within the ultradian timescale of multi-modal wearable sensor data. We hope to provide insight into the common latent states which occur, patterns in the time of day they occur at, and their relationship with self-reported exercise and mood.

\section{Methods}
\begin{figure*}
    \centering
    \includegraphics[width=1\linewidth]{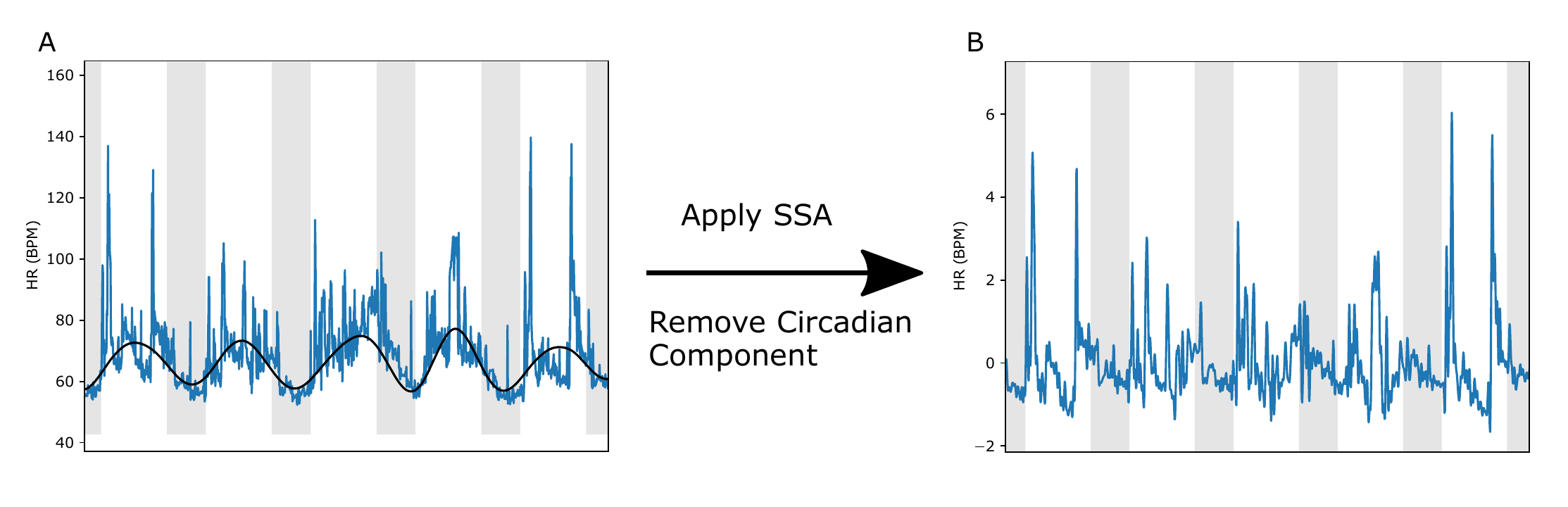}
    \caption{A) Shows an example of one sensor measurement, the circadian component of it (black line). Shaded areas indicate night time. To remove the circadian component and other lower or higher frequency noise we apply singular spectrum analysis (SSA) to decompose the signal. B) We then use Fourier analysis to identify the peak frequency of each component, then recombine components with peak frequencies between 48 and 18 cycles per day, giving us the signal on the right.}
    \label{fig:ssa}
\end{figure*}
\subsection{Data Collection}
We recruited 9 participants from within the staff and students of Newcastle University. Participants ranged in age from to 29 to 38 years old, 5 were female and 4 were male, and all were either students or professional office-based workers. 
We used three devices: a smartwatch - the Empatica EmbracePlus \cite{Gerboni2023} (measuring peripheral body temperature, electrodermal activity, and accelerometry), a continuous blood glucose monitor (CGM) - the Freestyle Libre 2 \cite{Fokkert2017}, and an ambulatory electrocardiogram (ECG) - the Bittium Faros 180  \cite{Bittium2019} (measuring heart rate and heart rate variability). Participants were given the option of wearing the ECG either with a single patch electrode, or in a two electrode configuration.
Participants were invited to the laboratory at the university to get set up with and initially put on the devices. First the researcher took written consent, then assisted them with applying the CGM and ECG, setting up the smartwatch, and then provided written instructions for the use and maintenance of all devices. All participants were asked to wear the sensors for two weeks and good compliance was observed.

\subsection{Calculating Metrics}
From the wearable sensors we calculated six measures - acceleration (Acc), temperature (Temp), electrodermal activity (EDA), continuous blood glucose (Glucose), heart rate (HR), and heart rate variability (HRV). Acceleration was quantified using the standard deviation of the acceleration, measured from the wrist. Temperature and electrodermal activity were also measured from the wrist using the smartwatch. For the electrodermal activity (the conductance of the skin) we used the skin conductance level - the slowly varying component of the raw measurement. Continuous blood glucose was measured using the CGM which provided an estimate of blood glucose every 15 minutes. Heart rate was calculated using R-R intervals derived from the ECG. Intervals outside of the physiological range (between 300 and 1200 milliseconds) were discarded. Heart rate variability was calculated as the standard deviation of the R-R intervals (SDRRI) over 5 minute epochs \cite{Shaffer2017}. 

\subsection{Isolating Ultradian Fluctuations}
All signals were first resampled to the mean of each five minute epoch - producing signals with a common sampling interval of five minutes. Missing data were replaced by linear interpolation for gaps shorter than 24 hours. If there were longer periods of missing data for any measure we removed chunks of the missing data across all measures in 24 hour blocks until any gap was less than 24 hours, and then used linear interpolation to replace any remaining missing data. This brought forward future data to narrow the gap while ensuring that all measures retained their original time stamp (if not their date stamp). Only one participant had missing data lasting more than 24 hours. 

As wearable sensor measures typically show fluctuations on a range of timescales (with a particularly strong circadian component) we wished to decompose each signal to isolate only those fluctuations occurring on an ultradian timescale (fluctuations with a period between 30 minutes and 12 hours). Univariate singular spectrum analysis (SSA), a non-parametric signal decomposition method \cite{Alessio2016}, was applied to decompose each signal into its periodic or quasi-periodic components. A \emph{window length} of 288 samples (24 hours) was selected because all components of interest had a period of between 0.5 and 24 hours. We first confirmed that each signal showed a strong circadian component - figure \ref{fig:ssa} (A) illustrates an example of the circadian component of heart rate, extracted by SSA. To isolate only the ultradian components, we used Fourier analysis to identify the peak frequency (presented as period for sake of interpretation) of each component, and then discarded any component with a peak period greater than 12 hours or less than 0.5 hours before taking the sum of the remaining components. This removed high frequency noise, the circadian component, and the baseline, so that the final signal represents ultradian fluctuations away from baseline. An example of this ultradian signal for heart rate is shown in figure \ref{fig:ssa} (B).
\subsection{Applying the HSMM}

The HSMM can be defined by the sequence of observations ($Y_{i,t}$, the observation in measure $i$ at time $t$), a state sequence ($C_{t}, C \in \{S_0 .. S_M\}$ the state at time $t$, where $M$ is the number of possible states), the distribution of observations associated with each state ($f_0(x) ..f_M(x)$, where each $f(x)$ is a multivariate distribution across all measurements), the state duration distributions ($ d_0(x) ..d_M(x)$), and the transition probability matrix $\Gamma$. 
The observation distributions ($f_0(x) ..f_M(x)$) are modelled as a multivariate Gaussian distribution (capturing the value of each signal while in the state) and the state duration distributions ($d_0(x) ..d_M(x)$) as a Poisson distribution. The transition matrix $\Gamma$, is an $M \times M$ matrix indicating the probability of transitioning to a state when in every other state. We used the pyhsmm Python package \cite{Johnso} to train the HSMM, which implements a Hierarchical Dirichlet Process Hidden semi-Markov Model \cite{Johnson2013}. We set the maximum state duration to 50 minutes (10 samples). Bayesian Information Criteria (BIC) was used to assess goodness-of-fit.  Equation \ref{eq:bic} shows how we calculated the BIC, where $M$ is the number of states, $LL$ is the log likelihood of the data given the model, $k$ is the number of parameters in the model, and $N$ is the number of samples the model has been trained on. 
\begin{equation}
    BIC = -2 * LL + k * log(N)
    \label{eq:bic}
\end{equation}
We set the maximum number of states to be 26 - this decision was based on an exploration of the BIC during training, after 26 states had been assigned the BIC had stabilised.
We interpret each state by looking at the parameters of their observation distributions - the mean ($\mu$) of each dimension of the multivariate Gaussian distribution. We also characterise the time distribution of each state by plotting a histogram of the number of samples spent in that state at each time of day. This indicates whether time spent in a given state is distributed throughout the day and night or concentrated to a specific time. 

\subsection{Measuring Physical Activity and Mood}
Self-reported physical activity and mood were recorded with ecological momentary assessment (EMA), using the Avicenna app \cite{Avicenna}. This prompted users 5 times a day, asking them if they had done exercise since last reporting, if they answered yes, they were then asked to specify the exact time span they were exercising. During the same prompt participants were also asked about their mood. This was in the form of two sliding scale ratings: one from ``Tired or Exhausted'' to ``Energised or Excited'' (Energy), another from ``Angry or Tense'' to ``Calm or Relaxed'' (Calmness). Scales ranged from -2 (negative) to 2 (positive), responses were normalised to the baseline of each participant by subtracting the mean response. Prompts were scheduled randomly between 10:00 and 21:00 (typical waking hours) using a uniform distribution.  

\section{Results}
\begin{figure*}
    \centering
    \includegraphics[width=1\linewidth]{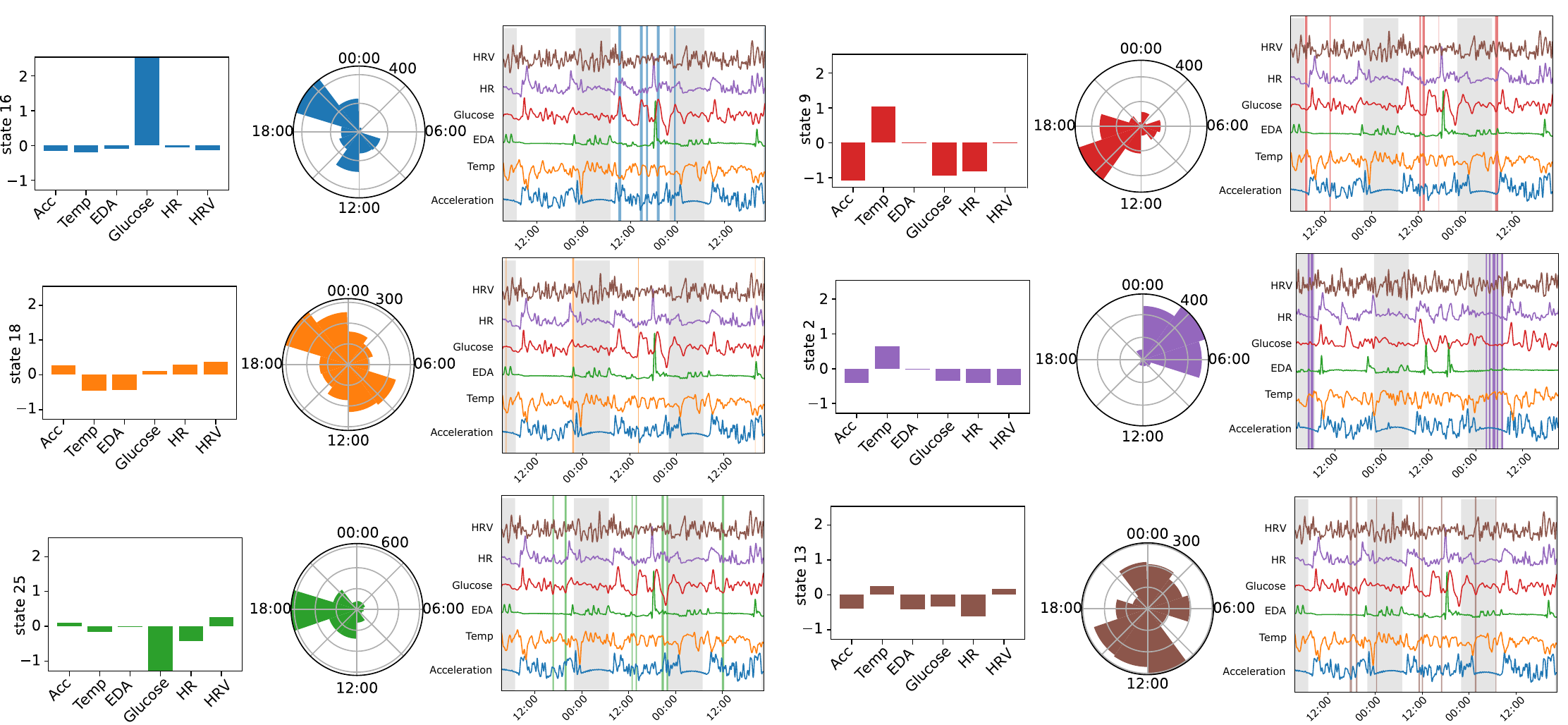}
    \caption{An illustration of the six most common states. For each state we show the mean of each metric as estimated by the HSMM ($\mu$ parameter of the distribution), the time distribution of each state, and an example trace of the 5 metrics with the time spent in the state highlighted.}
    \label{fig:examplestates}
\end{figure*}

\subsection{Common latent states}
\begin{figure}
    \centering
    \includegraphics[width=1\linewidth]{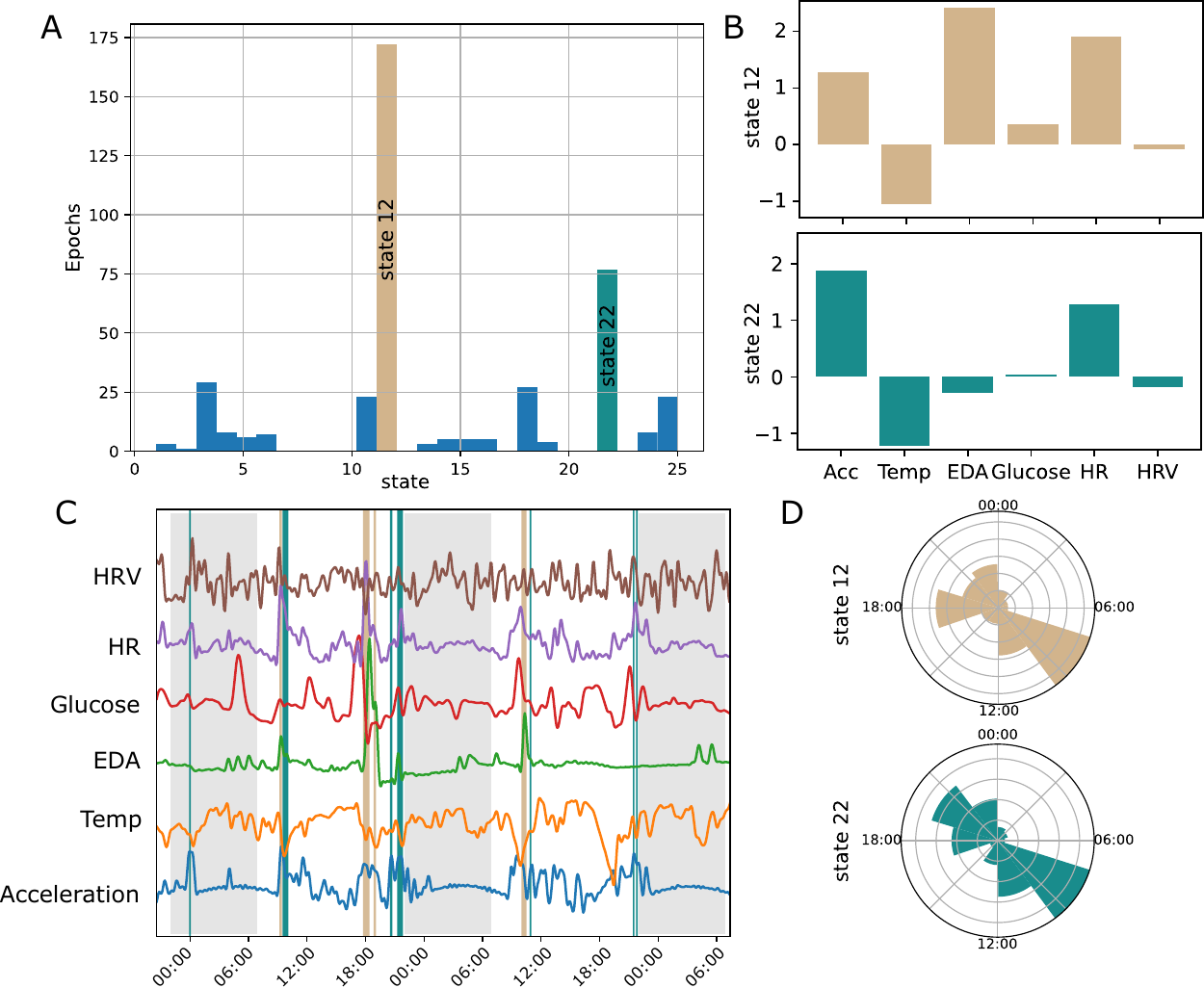}
    \caption{The HSMM states identified during physical activity. A) Shows the time (in 5 minute epochs) in each of the 26 states in the model. States 12 and 22, involve the most time spent doing physical activity. B) The HSMM estimates for the mean of states 12 and 22. C) An example of the signals recorded during states 12 and 22. D) The time distribution of states 12 and 22.}
    \label{fig:exercise}
\end{figure}
\begin{figure*}
    \centering
    \includegraphics[width=1\linewidth]{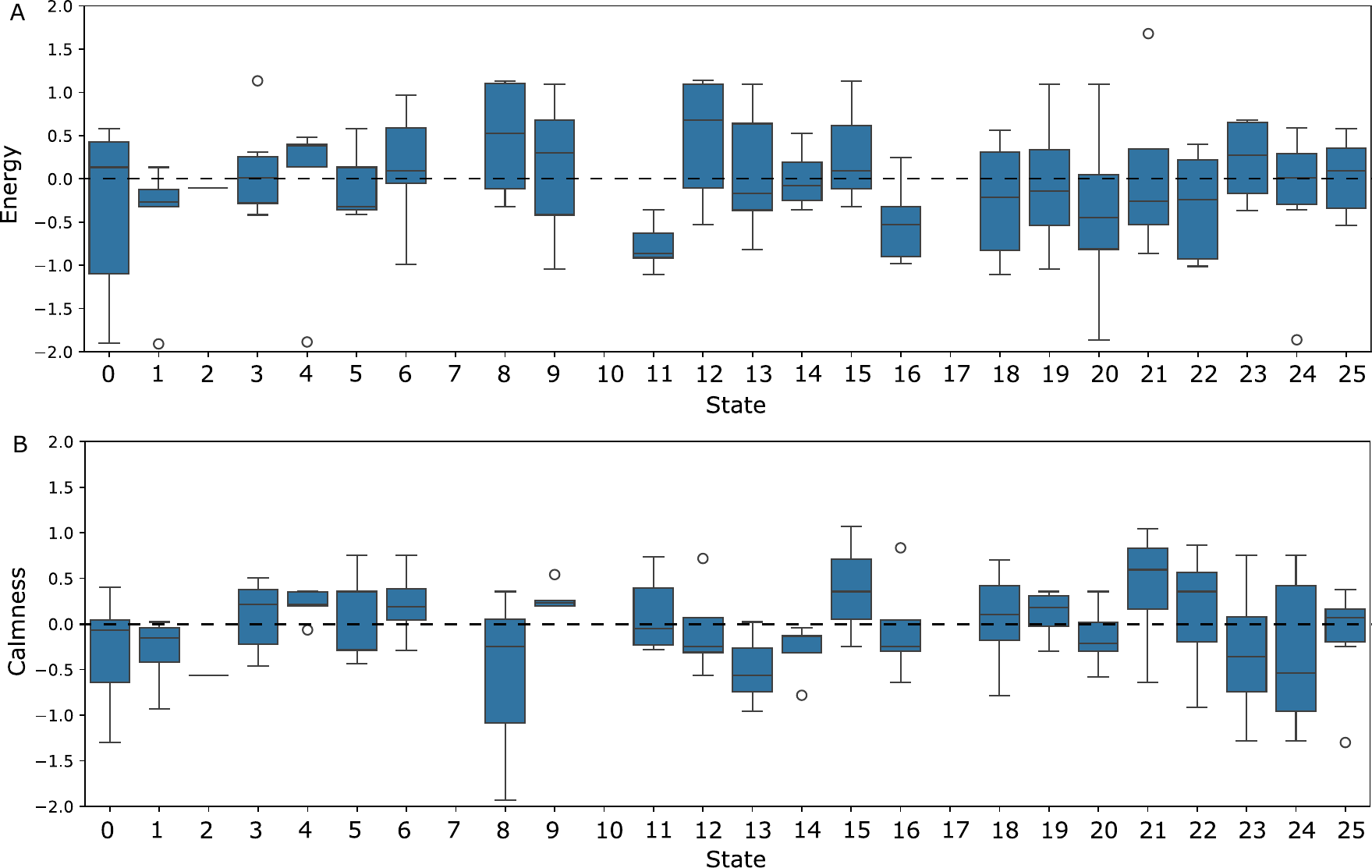}
    \caption{Relative mood ((A) energy, (B) calmness) assessed by ecological momentary assessment, during each of the states. Positive values indicate more energy or more calmness. Mood scores are normalised to the overall mood rating in that dimension for each participant. Some states, have no recorded mood for any participant. }
    \label{fig:mood}
\end{figure*}
The HSMM identified 26 latent states within the data. To illustrate the model we describe the 6 most used states, corresponding to those that participants spent the most time in. 
Figure \ref{fig:examplestates} shows the examples from the 6 most common states, showing the parameters (mean) of the multivariate Gaussian distribution associated with the observations of each state, the time distribution of each state indicating the time of day when they occur across all participants, and an example of the signals during each state. State 16 shows above baseline blood glucose but baseline levels of all other measures. It also occurs most frequently between 12:00 and 14:00 and between 19:00 and 21:00, corresponding to the time when the two largest meals are eaten. We can therefore interpret this as the glucose spike associated with eating.

Other states have less obvious interpretations but often show strong phase-locking to particular times of day and distinct combinations of observation parameters, indicating they may represent meaningful bio-behavioural states with potential relevance to health. For example state 9 occurs mostly in the mid-afternoon, and is characterised by below baseline movement, high skin temperature, low glucose, and low heart rate. This shows a pattern of measures similar to states which occur during sleep, such as state 2, and may indicate a period of mid-afternoon inactivity approaching sleep, commonly experienced by office workers. All states were also well represented across participants, in no case was a single participant responsible for the majority of time in a state. 
\subsection{Latent states associated with physical activity}
Physical activity is an important health behaviour, typically measured using time-use diaries or accelerometry. Here we show the potential of the HSMM approach to capture this important bio-behavioural state in an unsupervised manner. As participants noted the timing of their physical activity, we were able to describe which latent states they resided in during this time. Figure \ref{fig:exercise} shows the number of 5 minute physical activity epochs belonging to each state. Two states capture the majority of time spent in doing physical activity. State 12 is the most occupied state during self-reported physical activity, characterised by high acceleration, below baseline skin temperature, high EDA, baseline glucose, high heart rate, and baseline HRV. High heart rate and acceleration are key markers of physical activity, high EDA is also to be expected as a result of sweating, and low skin temperature could also be expected as exercise may take place outdoors. State 22 shows a similar pattern, the only clear difference is that here EDA is at a baseline level. This may indicate that this was exercise below the threshold for sweating, or that the EDA contacts were not optimal. Both states show similar time distributions, with peaks in the morning and the evening - plausible exercise times for professionals and students who spend the middle of the day working at home or in the office.
\subsection{Mood across latent states}
Two dimensions of mood were captured in 5 snapshots per day. Figure \ref{fig:mood} (A) shows the distribution of energy dimension ratings for each state and figure \ref{fig:mood} (B) shows the distribution of calmness dimension values for each state. The small sample size limits the conclusions we can draw from these - comparing each state and mood dimension would require 62 individual comparisons. However, states 11 and 16 have consistent ratings of tiredness, and states 9 and 21 have consistent ratings of calmness. We hope to expand this dataset so that we can confirm whether some states identified are associated with particular mood states. 
\section{Discussion}
\subsection{Main findings}
We found that the HSMM could identify latent states from ultradian fluctuations in the 6 measures. Many of these states showed a time-of-day preference consistent across participants, indicating that they may reflect bio-behavioural states that have strong time-of-day preferences as a result of chronobiology (e.g. sleep states) or sociological constraints (e.g. meal times, exercise times). 
We found that two states emerged that coincided with time spent exercising according to participants self-reports. The observation distributions of these states provided an indication of what exercise looked like physiologically - above baseline acceleration and heart rate, baseline glucose and HRV, and above baseline EDA (state 12) or baseline EDA (state 22).  Classification of exercise from wearable sensors is well demonstrated using supervised methods \cite{Kantoch}, and so it is not surprising that we were able to find states which strongly associated with exercise. However, demonstrating that this method identifies these clear bio-behavioural states gives us confidence that the less obvious states may also have relevance. 
\subsection{Comparison with previous work}
Much work has gone into applying machine learning - both supervised and unsupervised - to human activity recognition using wearable sensors, including HSMMs \cite{Duong2009}. At the same time, identifying the various chronobiological components of wearable sensor signals has also developed. SSA is an established method used to analyse chronobiological rhythms, and previous work has used it to isolate the circadian components of accelerometry data \cite{Fossion2017, Cui2023}. We have not been able to find any studies that have combined these approaches, recognising that identifying latent states in wearable sensor measurements should begin by isolating the chronobiological components of most interest. 
\subsection{Limitations}
Our sample size of 9 participants with a narrow age range, and similar sociodemographics is a major limitation of this study. Future work that incorporates a larger sample, and purposefully samples from a wider range of ages, occupations, and lifestyles will be important to further develop these methods. Age is known to influence our chronobiology, behaviour, and physiology \cite{Cornelissen2017}, while shift work and lifestyle or life events that shift our sleep-wake cycle may disrupt our chronobiological rhythms \cite{Cheng2021}. Accounting for this variety will entail further development of the methodology to recognise that some latent states will be present only for a subset of the population. However, this also raises the possibility of clustering the population using according to how often, and at what time, they enter particular bio-behavioural states. 
\subsection{Potential for further insights}
The HSMM also models the probability of transitioning from one state to another. This probability matrix could provide further insight into the relationships between states - for example we could ask the question \emph{does state x regularly follow state y?} This may provide an additional layer of understanding and potential for novel interventions - for example if we know that there is a high probability to transition to state x when in state y, and we wished to avoid state x, we may then design an intervention that reduces time spent in state y. 

\section{Conclusion}
Here we have presented a novel approach towards developing a data-driven understanding of multi-modal wearable sensor measurements. We have shown that by extracting only the ultradian rhythm components of each signal, we can identify latent states using a HSMM which can provide insight into the bio-behavioural state of the wearer. We hope this will lead to new methodologies for measuring health behaviours, allowing researchers to utilise multi-modal data to quantify how participants spend their time - physiologically and behaviourally. We also envision that this approach could be used within healthcare to identify bio-behavioural states associated with specific diseases. Future work will explore how time spent in particular latent states relates to health outcomes such as well-being and stress management and further data collection should provide confirmation of any association with mood and explore whether latent states can map to more complex self-reported behaviours.

\section*{Acknowledgment}

\bibliographystyle{IEEEtran}
\bibliography{IEEEabrv, ref}

\end{document}